# Optics-less Sensors for Localization of Radiation Sources

**H.J. Caulfield[1], L. P. Yaroslavsky[2], Ch. Goerzen[2], S. Umansky[2]**

[1] *Physics Department, Fisk University*
*1000 17th St., Nashville, TN 37298, USA*
[2] *Dept. Physical Electronics. Faculty of Engineering. Tel Aviv University.*
*Tel Aviv, Ramat Aviv 69978, Israel*

A new family of radiation sensors is introduced which do not require any optics. The sensors consist of arrays of elementary sub-sensors with natural cosine-law or similar angular sensitivity supplemented with a signal processing unit that computes optimal statistical estimations of source parameters. We show, both theoretically and by computer simulation, that such sensors are capable of accurate localization and intensity estimation of a given number of radiation sources and of imaging of a given number of sources in known positions. The accuracy is found to be dependent only on the sub-sensors' noise level, on the number of sub-sensors and on the spacing between radiation sources.

One can treat images as data that indicate locations in space and intensities of sources of radiation. Conventional optical imaging systems use photo-sensitive plane arrays of sub-sensors coupled with focused optics that form a map of the environment on this image plane. The optics carry this out at the speed of light, but come with some disadvantages. Because of the law of diffraction, accurate mapping requires large lens sizes and complex optical systems. Also, lenses limit the field of view and are only available within a limited range of the electromagnetic spectrum. The ever-decreasing cost of computing makes it possible to make imaging devices smaller, by replacing optical and mechanical components with computation. This motivates a search for optics-less computational imaging devices.

The methods presented in the paper draw heavily on estimation theory. Since Fisher's publication on Maximum Likelihood (ML) estimation ([1]), many articles have applied this theory to specific problems. Van Tree's textbook [2] gives a good overview of the statistical methods used in this article. Works in image restoration such as that by Lucy [3] use a ML or maximum a-posteriori probability reconstruction of the image. Recently, researchers have begun to study artificial compound eyes (see for instance, Refs. [4,6-11]). In particular, Ref. [4] uses information theory to predict performance limits of compound eyes of insects and Ref. [11] uses estimation of sources for sub-pixel estimation of source angles in an approach similar to our own. In general, the use of optimal statistical estimations for measuring physical parameter of objects using multiple sensors is very common in sensorics ([12-14]).

This paper applies the ML method to perform imaging tasks by means of imaging devices consisting of a set of sub-sensors with no devices added to reshape the sub-sensor's natural angular sensitivity to radiation. In this respect sensors discussed in this paper are significantly different from artificial compound eyes. We will show that, in spite of relatively wide angular sensitivity of the sub-sensors, our sensors are capable of quite accurate localization and imaging of radiation sources provided the number of sub-sensors and their signal-to-noise ratios are sufficiently high. Using commonly accepted terminology (see, for instance, [5]), we call the class of sensors introduced in the paper "Optics-less Smart" (OLS) sensors.

OLS sensors have extremely simple physical design. Examples of possible designs for the OLS sensors are sketched in Fig. 1. In the sensor arrays shown in Fig. 1, a) and b), the elementary sub-sensors are placed on the outer surface of a sphere. Such sensors are capable of localizing radiation sources throughout the entire solid angle. Fig. 1, c) shows an alternative design, an array of elementary sub-sensors on a flat surface. Because rays from distant sources arrive at the flat sensor from the same angles, this sensor is capable of measuring coordinates and intensities only of sources that are sufficiently close to the sensor. Any intermediate design consisting of sub-sensors on curved convex or concave surfaces is also possible. This paper is mostly concerned with the case of the spherical OLS sensor.

The operational principle of OLS sensors can be explained using the special case of estimating the locations of a known number of $\boldsymbol{K}$ point radiation sources by an array of $\boldsymbol{N}$ elementary sub-sensors. Let $\boldsymbol{I[k]}$ be the intensity of the $\boldsymbol{k}$-th source, $\boldsymbol{\theta_{SRC}[k]}$ be the directional angle to the $\boldsymbol{k}$-th source, $\boldsymbol{\theta_{SENS}[n]}$ be the angle of the surface normal of $\boldsymbol{n}$-th elementary sub-sensor and $\boldsymbol{s[n]}$ be its response. Assume also that this response contains additive signal-independent sensor noise $\boldsymbol{v[n]}$. Then:

$$\boldsymbol{s[n] = \sum_{k=1}^{K} I[k] \,\mathrm{costr}\left(\left|\theta_{SRC}[k] - \theta_{SENS}[n]\right|\right) + v[n]} \qquad (1)$$

is a model for the response of a sensor with cosine-law angular sensitivity, where $\boldsymbol{\mathrm{costr}(\gamma) \equiv \cos(\gamma)}$ for $\boldsymbol{|\gamma| \le \pi/2}$ and 0 otherwise. The sensor's signal processor operates on output signals $\{s[n]\}$ of all $N$ elementary sub-sensors and generates estimates $\{\hat{I}[k]\}$ and $\{\hat{\theta}_{SRC}[k]\}$ of intensities and directional angles of the radiation sources. In view of the statistical nature of sensor noise, the Bayesian approach to optimal statistical estimations can be applied. In order to facilitate further explanations, we opt to use ML estimation. Any other optimal statistical estimators, ([2]) may also be considered. Changing the type of the estimator requires only the corresponding re-programming of the sensor signal processing unit.

Assuming that elementary sub-sensor noise components are statistically independent and normally distributed with variance $\sigma^2$, the ML estimator is:

$$\left\{\hat{I}[k],\hat{\theta}_{SRC}[k]\right\}=\underset{\{I[k],\theta_{SRC}[k]\}}{\mathbf{argmin}}\left\{\sum_{n=1}^{N}\left(s[n]-\sum_{k=1}^{K}I[k]\mathbf{cos\,tr}\left(\left|\theta_{SRC}[k]-\theta_{SENS}[n]\right|\right)\right)^2\right\} \quad (2)$$

For a single radiation source, the computational complexity of the estimator is of order $N^{3/2}$ and the computations can be implemented in a simple hardware. Solving this equation for multiple sources means finding a minimum in $2K$-dimensional space. In this case, the computational complexity of the estimator grows exponentially with the number of sources in the naïve solution. There are ways to reduce the optimization computational complexity to polynomial, however they are out of the scope of the paper.

For multiple radiation sources, OLS sensors can be used in following modes:

- **"General localization" mode** for localization and intensity estimation of a known number of radiation sources.
- **"Imaging" mode** for estimation of intensities of a given number of radiation sources in the given locations, for instance, on a regular grid.

The sensors were tested in these two modes by numerical simulation using, for solving Eq. 2, the multi-start global optimization method with pseudo-random initial guesses ([15]) and Matlab's quasi-Newton method for finding local optima. The work of OLS sensors in "imaging" mode is illustrated in Fig. 2 for a model of the OLS spherical sensor consisting of 300 elementary sub-sensors set to estimate intensities of sources arranged in a form the abbreviation "OLSS", which stands for "Optics-less smart sensors." Note that operation in imaging mode can also be regarded as deblurring of raw images on the output of the sub-sensors (see Fig. 2, center).

Performance of OLS sensors is characterized, in first order approximation, by estimation variances. Theoretical analysis shows that, for sub-sensors with signal independent Gaussian additive noise, estimation errors have a normal distribution with mean of zero and standard deviation given by the Cramér-Rao lower bound (CRLB) [2]). We have derived for the 1D model (Fig. 1, a) of the spherical sensor of $N$ of elementary sub-sensors and a single source, that the CRLB for standard deviations of estimates $\hat{I}$ and $\hat{\theta}_{SRC}$ of source intensity and direction gives:

$$\sqrt{\mathrm{var}\left\{\hat{\theta}_{SRC}\right\}} \geq \frac{2\sigma}{\sqrt{NI}}; \quad \sqrt{\mathrm{var}\left\{\hat{I}\right\}} \geq \frac{2\sigma}{\sqrt{N}} \quad . \tag{3}$$

The covariance terms in the CRLB are all zeros. In the case of a single source, the inequalities in Eq. 3 may be replaced by equalities.

Results of numerical runs in our Monte-Carlo simulations, plotted in Fig. 3, confirm that the simulated sensor's performance matches the CRLB. However for multiple sources, the error may be significantly higher, especially if neighboring sources are very close to one another. For the case of two sources, we derived, by means of a simple argument based on the single-source CRLB, that standard deviations of estimates $\hat{I}$ and $\hat{\theta}_{SRC}$ of source intensity and direction are:

$$\sqrt{\mathrm{var}\left\{\hat{\theta}_{SRC}[k]\right\}} = \frac{2\sigma}{I[k]\sqrt{N \cdot \frac{\Delta\theta_{SRC}}{\pi}}}; \quad \text{given} \quad \frac{N \cdot \Delta\theta_{SRC}}{\pi} > 1; k = 1,2 \tag{4}$$

where $\Delta\theta_{SRC}$ is the angular difference between sources. In the case where the angular separation between sources is smaller then the angular separation between neighboring sub-sensors ( $N \cdot \Delta\theta_{SRC} / \pi \leq 1$), the estimator's performance is no better than random guessing. Eq. 4 suggests that two sources can be resolved only if the angular distance between the sources is greater or equal to the angular distance between sub-sensors. This finding is supported by numerical runs, shown in Figs. 4, a) and b). However, further studies using angular sensitivity functions similar to but not identical to cosine show that with sub-sensors with these functions much higher angular resolution may be achieved.

Eq. 4 is theoretically valid only for the two-source case, but numerical results for cases with more than two sources show that regardless of the number of equally-spaced sources, the average estimation error for all the sources is equal to the error predicted for the 2-source problem. Although the situation for multiple sources requires further study in order to arrive at a good theoretical understanding, the simulation results make evident that the described OLS sensors are capable of locating radiation sources and evaluating their intensities in spite of quite poor angular selectivity of their elementary sub-sensors.

As always, there is a tradeoff between good and bad features of the OLS sensors that have to enter into a system design. The advantages include the following:

- No optics are needed, making this type of sensor applicable to virtually any type of radiation and to any wavelength
- The field of view (of spherical sensors) is unlimited
- The sensor's resolving power is determined ultimately by the sub-sensor's signal-to-noise ratio, and not by diffraction-related limits

The cost for these advantages is the high computational complexity, especially when good imaging properties for multiple sources are required. However, the inexorable march of Moore's law makes this problem less restrictive each year. Furthermore, the computations lend themselves to high-concurrency computation, so the computational aspects are not expected to hinder usage.

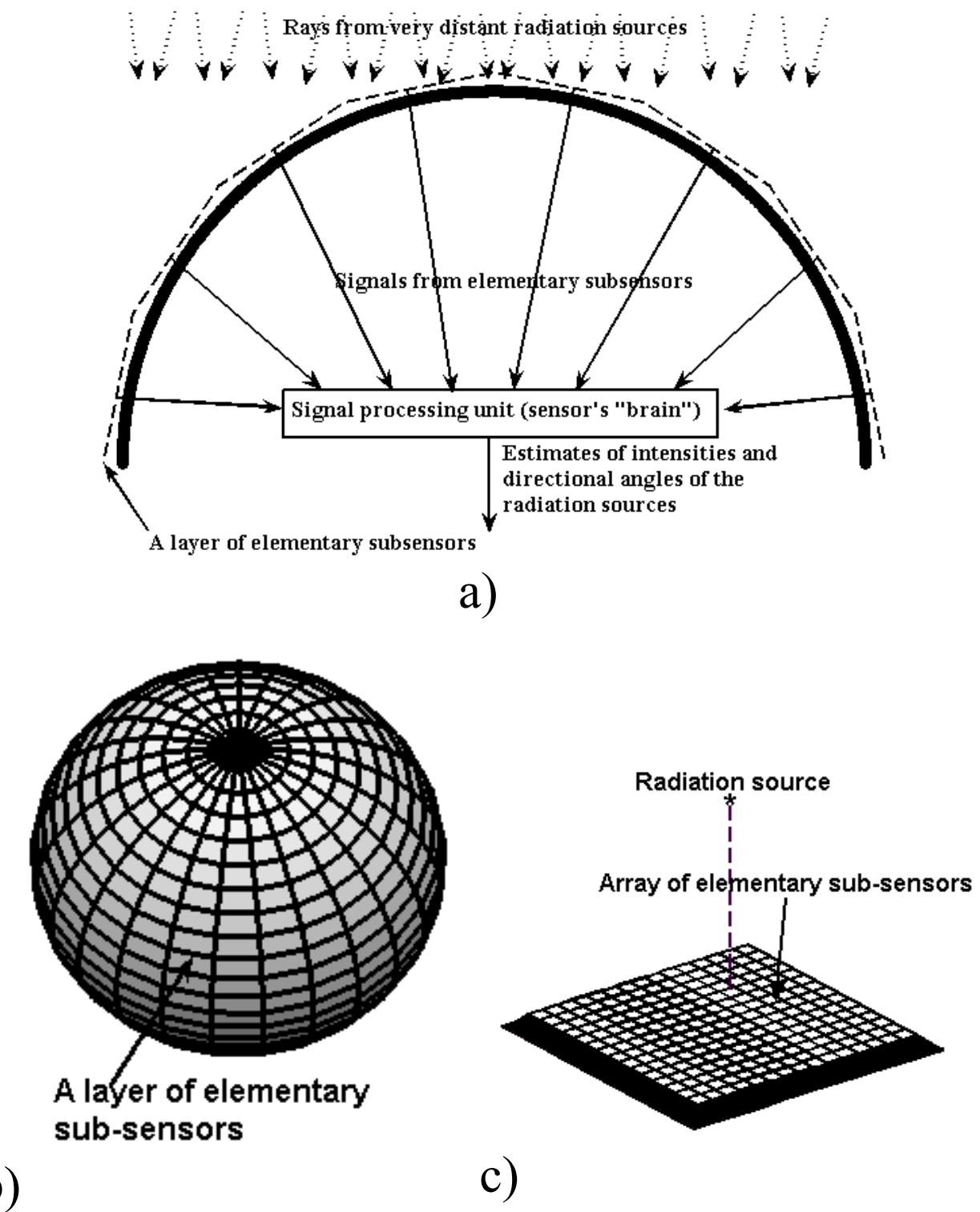

Fig. 1: Schematic diagram and examples of possible designs of OLS radiation sensors for very distant sources (a, b) and for sources in their close proximity (c)

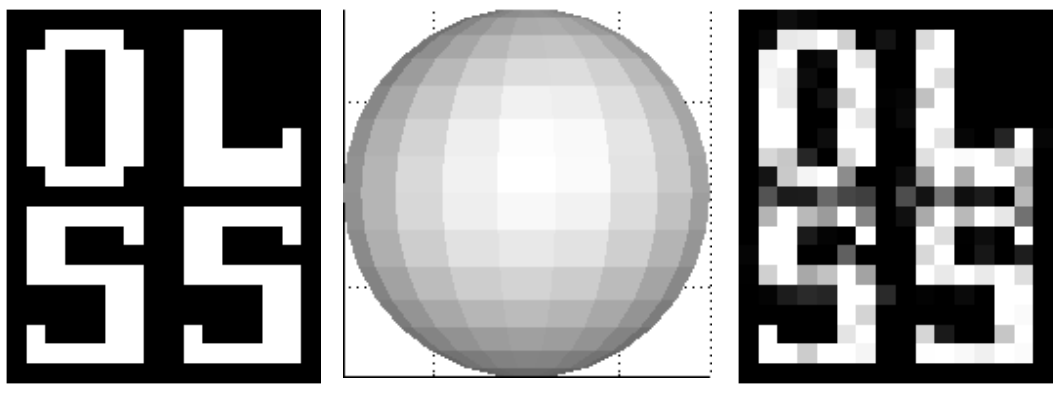

Fig. 2: OLS spherical sensor in the "imaging" mode: the original image is an array of sources that form characters "OLSS"(left), the blurred image is shown on the surface of the sphere (center), and the estimates of source intensities (right). The sensor consisted of 15x20 = 300 sub-sensors arranged within spatial angles $\pm\pi$ longitude and $\pm\pi/2.05$ latitude. The array of simulated radiation sources consisted of 19x16=304 sources with known directional angles within spatial angles $\pm\pi/2$ longitude and $\pm\pi/3$ latitude. Each sub-sensor had a noise standard deviation of 0.01, and source intensities were 0 (dark) or 1 (bright). Standard deviation of estimation errors was 0.0640.

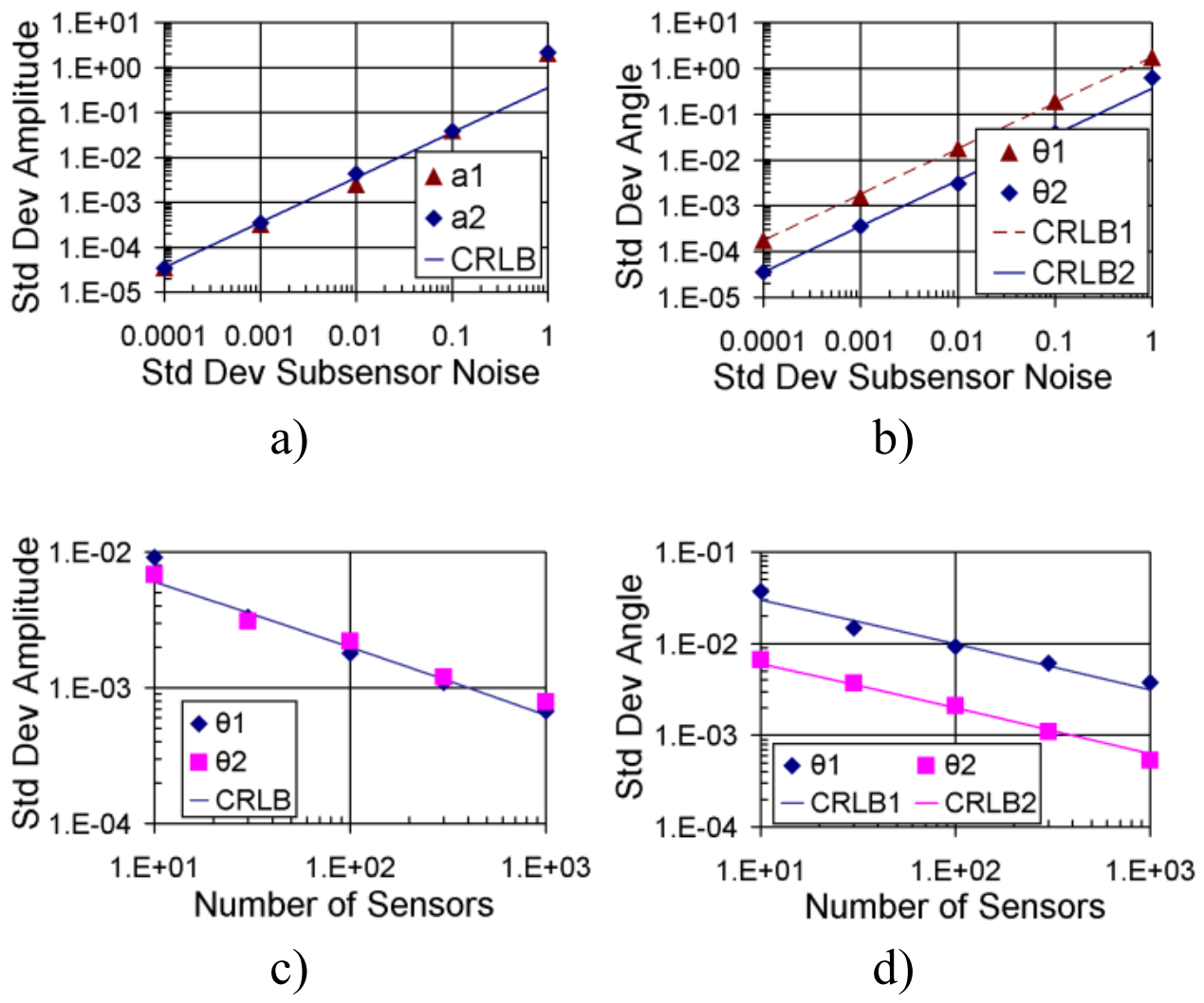

Fig. 3: Angular and intensity error standard deviations and Cramér-Rao lower bound (Eq. 3) for 1-D model of a circular OLS sensor. (a,b) show estimation error as functions of sub-sensors noise with number of sub-sensors fixed at 30; (c,d) show estimation error as functions of the number of sub-sensors with sub-sensor noise standard deviation fixed at 0.01. The two radiation sources with directional angles "$\theta_1$" and "$\theta_2$" are separated by π radians, and so may be considered as two separate runs of a single source, and in all runs first source intensity $a_1$ equals 1.0 and second source intensity $a_2$ equals 0.2.

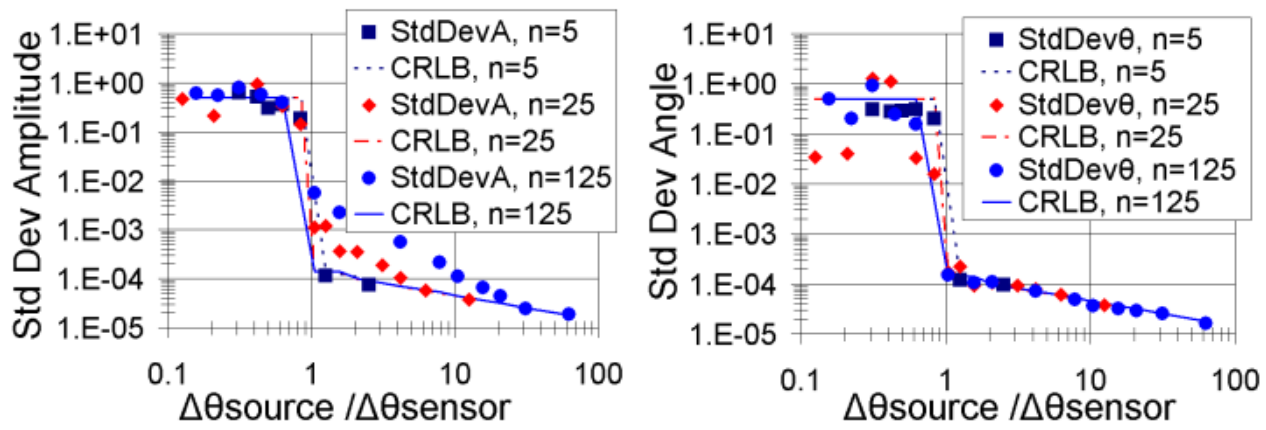

a) b)

Fig. 4: Experimental data and theoretical curves (Eq. 4) for error in estimating parameters for two sources. a) shows average error standard deviation in amplitude while b) shows error standard deviation in angle. In all runs, source amplitudes both equal 1.0, and sub-sensor noise standard deviation is always $10^{-4}$.